\documentclass[a4paper,12pt,reqno,showkeys,superscriptaddress,nofootinbib]{revtex4}
\usepackage[centertags]{amsmath}
\usepackage{amsfonts}
\usepackage{amssymb}
\usepackage{amsthm}
\usepackage{newlfont}
\usepackage{stmaryrd}
\usepackage{mathrsfs}
\usepackage{mathtools}
\usepackage{euscript}
\usepackage{graphicx}
\usepackage{enumerate}
\usepackage{todonotes}

\usepackage{dblfloatfix}
\usepackage{color}

\usepackage{tikz}
\usepackage{pgf}
\usetikzlibrary{positioning,fit,calc}
\usetikzlibrary{arrows,automata}
\usepackage{wrapfig}
\usepackage{subfigure}
\usepackage{amscd}
\usepackage{hyperref}
\usepackage{cancel}

\theoremstyle{plain}

\theoremstyle{definition}

\theoremstyle{remark}

\numberwithin{equation}{section}

\let\al=\alpha \let\be=\beta  
\let\ve=\varepsilon

\newcommand{\opunit}{\text{1}\kern-0.22em\text{l}}

\DeclareMathAlphabet{\mathpzc}{OT1}{pzc}{m}{it}

\newcommand{\id}{\textrm{d}}

\usepackage{floatrow}
\usepackage{caption}

\begin{document}
\title{A Nernst heat theorem for nonequilibrium jump processes}
\author{Faezeh Khodabandehlou}
	\affiliation{Instituut voor Theoretische Fysica, KU Leuven, Belgium}
		\email{faezeh.khodabandehlou@kuleuven.be}
\author{Christian Maes}
	\affiliation{Instituut voor Theoretische Fysica, KU Leuven, Belgium}	\author{Karel Neto\v{c}n\'{y}}
\affiliation{Institute of Physics, Czech Academy of Sciences, Prague, Czech Republic}
\keywords{nonequilibrium jump process, excess heat, Nernst postulate}

\begin{abstract}
We discuss via general arguments and examples when and why the steady nonequilibrium heat capacity vanishes with temperature.  The framework is the one of Markov jump processes on finite connected graphs where the condition of local detailed balance allows to identify the heat fluxes, and where the discreteness more easily enables sufficient nondegeneracy of the stationary distribution at absolute zero, as under equilibrium.  However, for the nonequilibrium extension of the Third Law, a dynamic condition is needed as well:  the low-temperature dynamical activity and accessibility of the dominant state must remain sufficiently high so that relaxation times  do not start to dramatically differ between different initial states.  It suffices in fact that the relaxation times do not exceed the dissipation time.  
\end{abstract}
\maketitle

\section{Introduction} 
The Third Law of Thermodynamics is a cornerstone of physical and chemical thermodynamics \cite{cal,fer}.  It sets and constrains the scene of low-temperature phenomena, at least where it concerns thermal properties of equilibrium systems.  To witness the importance, the quantum revolution was much inspired by the difficulty to account in classical physics for the low-temperature behavior of specific heats \cite{kle}.  The Nernst postulate deals explicitly with these questions, arguing that,  at zero temperature, the equilibrium entropy becomes a constant independent of any external parameter, \cite{ner}. Combined with Boltzmann's interpretation of entropy, Planck's version states that this entropy constant is zero, \cite{pla}.  It provides a fixed reference point for the absolute entropy of any substance at any temperature. Further conditions imply that the heat capacity must vanish with temperature \cite{cal,mas} and the state of absolute zero is physically unattainable via a finite number of {\it equilibrium} steps. However, in contrast with the First and Second Law, the Third Law is not always obeyed by substances and corresponding Nernst theorems, therefore, require nontrivial conditions, \cite{lie1,lie2}.\\

Since a long time, it has been considered important to develop calorimetry and to find a similar general structure  in the study of thermodynamic processes connecting
steady states of nonequilibrium systems, \cite{pri,oon,kom2}. The present work elaborates on such a general and nonperturbative extension of the Third Law, valid at an arbitrary distance from equilibrium and operationally formulated using a physically significant quasipotential from which heat capacities and latent heats can be defined.  We explain a dynamical condition for those heats to vanish at absolute zero, requiring residual dynamical activity, which, when violated, enables the possibility of a zero--temperature phase transition.\\
As is well-known, general results on thermal properties of nonequilibrium systems (steady or transient for that matter) are very few, with the positivity of the entropy production, steady and transient fluctuation theorems, thermodynamic and kinetic uncertainty relations, and fluctuation-response relations being the most important ones; even there however the zero-temperature limit is seldom discussed (see \cite{sor} for an interesting exception).\\

The point is not at all to formally extend the definition of entropy \cite{scr}, and we do not need a Clausius heat theorem out-of-equilibrium.  Rather, we wish to obtain quantitative measures of the heat released or absorbed during a quasistatic relaxation to a (new) nonequilibrium condition {\it because of} (small) changes in control parameters.  That also applies to small systems and need not be restricted to the context of linear irreversible thermodynamics or to close-to-equilibrium processes. Far-from-equilibrium examples within electronics are micro- and nano-scale devices as digital logic circuits with a rich collection of nonlinear dissipative effects due to quantum tunneling, current leakage, etc.~\cite{pop}. Other examples include driven hopping dynamics on a lattice in contact with an electron- or photon-bath, where the driving can be electromagnetic or optical, \cite{dls,qa}.  As a third class of examples and motivation, we can turn to thermochemistry. Under detailed balance, the energy exchanges during chemical reactions are measured with the methods of calorimetry, and molar heat capacities inform about the landscape of thermodynamic potentials.  Thus, at (very) low ambient temperatures, cryogenic chemistry mostly focuses on the ability of molecules to lower the total energy. We remember that Nernst got the 1920 Nobel Prize in chemistry ``in recognition of his work in thermochemistry.''  However, under nonequilibrium conditions where detailed balance is violated, kinetic aspects will matter greatly and it is essential to understand the constraints set by an extended Third Law.  While, in the above examples, the driving can be time-dependent (like when applying an AC-field or periodic feeding), here we imagine time-scales where the external field can be taken almost constant.  The extension to truly time-dependent external driving is conceptually not different though, and we give an example of random switching in Section \ref{heac}.\\

In driven systems, the energy exchange always runs against a dissipation background. In the stationary situation, work
done by nonconservative forces is constantly dissipated
as (Joule) heat; see the cartoon in Fig.\ref{exh} (left). Or, when in contact with different chemical reservoirs such as in chemostats and bioreactors, particle currents are maintained between external sources and sinks at some temperature.  Changing an external parameter, such as that ambient temperature, creates excess heat in the relaxation to the new steady condition. Then, our interest may go to that excess part coming from changes in the temperature and/or
other parameters defining the system dynamics \cite{HS,GP,cla}.
However, how the total heat exchange is split into the ``background'' and the ``excess'' is not unique and various schemes exist exhibiting different features. 
One such a scheme is known as the Hatano-Sasa approach \cite{HS}, where the total heat is decomposed into the ``house-keeping'' and the ``excess'' components in such a way that the excess heat exactly satisfies the (extended) Clausius heat theorem. Other variants of this approach can be found in \cite{ber,cla,GP,st}. Yet, measuring that excess heat is not obvious.  Another scheme, which goes back to the original proposal by Oono and Paniconi \cite{oon}, considers the excess heat defined by subtracting the time-integrated steady state heat flux \cite{kom2}, sometimes called the ``dynamical component'' of heat \cite{wang}. While such an excess heat does not satisfy the Clausius relation unless close to equilibrium, it enjoys the important properties of being both ``geometrical'' (= dependent only on the thermodynamic trajectory) and ``reversible'' (= antisymmetric with respect to protocol-reversal) in the quasistatic limit. The mathematical analysis of that limit is obtained in \cite{arxiv}.  Here we take the result in terms of the so-called     quasipotential and since that is also operationally accessible (e.g., by an AC-measurement with low-frequency temperature variations), see e.g. the cartoon of Fig.~\ref{exh} (right), that becomes a preferential option for the calorimetric analysis of the present paper. It allows a natural definition of the nonequilibrium heat capacity, \cite{eu,jir,calo}. The main result then is that for a large and physically motivated class of nonequilibrium systems modeled as Markov jump processes, that heat capacity vanishes at absolute zero.  We explain the context, significance, and main ideas in the present paper, also adding simple examples.

In the next section, we review the main elements of nonequilibrium calorimetry.  Section \ref{mjp} describes the framework of Markov jump processes and how calorimetry is set up in the absence of (global) detailed balance.\\  
Section \ref{nerp} is devoted to stating and explaining the extended Nernst postulate. The simplest sufficient condition for the validity of the extended Third Law is that relaxation times remain smaller than the dissipation time at low temperatures. Various illustrations are added in Section \ref{exs}, showing that our conditions are natural and not just sufficient.  Sections \ref{qai} and \ref{sig} provide extra background and discussion for better interpretation and significance of the results.  The conclusion section \ref{con} gives a short summary.\\

We note that the present paper focuses on the underlying ideas and intuition.  We not only skip the detailed analysis of the quasistatic limit, but we also present the result only for the heat capacity. In all it makes an essential complement to the mathematical proof which can be found in \cite{arxiv}.

\section{Nonequilibrium calorimetry}
We collect in the present Section a review of the general thermal physics that leads to the notion of a steady nonequilibrium heat capacity.  As such and independent of a specific mathematical setup or model, it has not appeared before but the main ingredients are also contained in the papers \cite{eu,jir,calo,ner}.  We note that other definitions of nonequilibrium heat capacity have been proposed, see e.g.~\cite{man,subas,dls}, such as the temperature-derivative of the mean energy on which we comment briefly in Section \ref{eqc}.

\subsection{Thermal response}
\begin{figure}[H]
\centering
\includegraphics[scale=0.6]{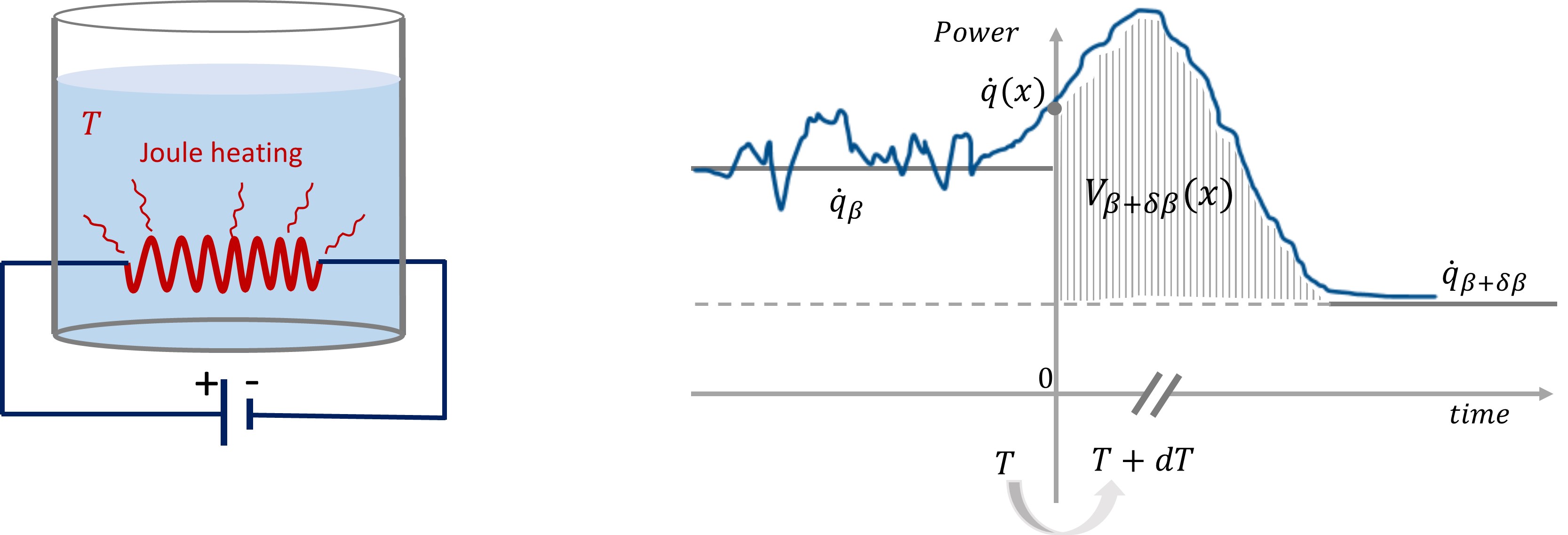}
\caption{\small{left: Changing the temperature $T$ of the bath changes the dissipated power in Joule heating. right: In equilibrium, the mean dissipated power ${\dot{q}}_\beta = {\dot{q}}_{\beta+\id \beta} =0$ vanishes, but under driving, they may take different positive values. The dissipated power is variable and fluctuates around its steady value ${\dot{q}}(\beta)$, as indicated for negative times. At time zero, a condition $x$ is obtained for the relaxation of the expected heat flux to its stationary value, indicated for positive times. The shaded area gives the meaning of the quasipotential.  }}\label{exh}
\end{figure}

Consider an open system that is in a steady nonequilibrium condition, maintaining stationary currents and a strictly positive mean dissipative power.  A simple example is depicted as a cartoon in the left part of Fig.~\ref{exh}: a current-carrying resistor creates Joule heating of the surrounding liquid.   It stands of course for many other examples, including systems in nonlinear regimes where the resistance depends in complicated ways on the bath temperature $T$ or where the voltage is a nonlinear function of the current.  At any rate, if the bath is sufficiently large  and the coupling with the system is weak, the temperature $T$ can be considered constant and the bath is and remains in thermal equilibrium.  Putting $\beta=T^{-1}$, we often move between the two notations for (inverse) temperature.   
Before time zero, there is mean dissipated power $\dot q_\beta\geq 0$.  Imagine next that very slowly (in the so-called quasistatic limit), the temperature of the bath is changed to a new value.  Such a procedure requires small steps in temperature, at a rate that is much slower than the inverse relaxation times.  To keep it simple, we consider in Fig.~\ref{exh} (right) only one step in temperature: at time $t=0$ the inverse temperature is changed $\beta+\id \beta$ and kept fixed at $\beta+\id \beta$.  Suppose the state is $x$ at time zero.  After a possibly long time, the system reaches a new steady nonequilibrium condition, now with power $\dot{q}_{\beta+\id \beta}\geq 0$, independent of $x$.  The shaded area in Fig.~\ref{exh} is the integral of a difference in power, and is given by the so-called quasipotential
\begin{equation}\label{qpt}
V(x,\beta+\id \beta) = \int_0^\infty \id t\,\left[{\dot{q}}_t(x, \beta+\id \beta) - {\dot{q}}_{\beta+\id \beta}\right]
\end{equation}
where ${\dot{q}}_t(x, \beta+\id \beta)$ is the expected dissipated power at time $t>0$ when starting (at time zero) from state $x$.  Note that we assume here that, at least for all finite $\beta$ and relevant conditions $x$, ${\dot{q}}_t(x, \beta+\id \beta) \rightarrow {\dot{q}}_{\beta+\id \beta}$ sufficiently fast as $t\uparrow \infty$ so that the integral in \eqref{qpt} is converging.\\

The quasipotential \eqref{qpt} measures an excess heat. There is obviously important literature on such excesses; we already mentioned \cite{oon,kom2,wang}, but there is also the notion of path-dependent heat and work, as used by many people studying nonequilibrium fluctuations, including e.g. \cite{ken, 2law}.\\
Remember now that $x$ in \eqref{qpt} and in Fig.~\ref{exh} (right) is a state which is statistically selected from the nonequilibrium ensemble which describes the static fluctuations at time zero when the system is still coupled with the thermal bath at  temperature $T$.  
The shaded area \eqref{qpt} has an average (over $x$) $\langle V(x,\beta+\id \beta)\rangle_\beta = O(\id \beta)$ of order $\id \beta$, since clearly $\langle V(x,\beta) \rangle_\beta =0$ from \eqref{qpt}. 
The (nonequilibrium) heat capacity is therefore mathematically and physically well-defined as
\begin{equation}\label{hcao}
C(T) = \beta^2\, \frac{\langle V(x,\beta+\id \beta)\rangle_\beta}{\id \beta} = \beta^2 \,\big\langle\, \frac{\id  V}{\id \beta}(x,\beta)\, \big\rangle_\beta
\end{equation}
measuring the extra heat {\it toward} the system per temperature change while we fix possible other external controls.


\subsection{Fluctuation expression}
The formula \eqref{hcao} for the heat capacity has an average over a steady nonequilibrium ensemble.  It is determined by the time-invariant probabilities, denoted by
$\rho (x) = \rho _\beta(x)$ where we often ignore the temperature-dependence in the notation, to find the system in state $x$ when the equilibrium bath (to which it is weakly coupled) is kept at inverse temperature $\beta$. Since $\langle V(x,\beta)\rangle_\beta=\sum_x \rho _\beta(x)V(x,\beta) =0$, we can continue from \eqref{hcao} to write
\begin{equation}\label{fluca}
C(\beta) = \beta^2 \sum_x \rho _\beta(x) \frac{\id  V}{\id \beta}(x,\beta) = -\beta^2\sum_x  \frac{\id  \rho _\beta}{\id \beta}(x) \,V(x,\beta)
\end{equation}
where for convenience we assumed a finite number of possible states $x$. (Otherwise, we would write an integral with $\rho (x)$ as probability density, or even more general expressions.) 
As a consequence, after a last rewriting of \eqref{fluca},
\begin{equation}\label{fluc}
C(\beta) = -\beta^2\big\langle \frac{\id \log \rho _\beta(x)}{\id \beta} \,\,V(x,\beta)\big\rangle_\beta
\end{equation}
which is a correlation function between the quasipotential $V(x,\beta)$, related to excess heat, and the temperature-dependence $\frac{\id}{\id \beta} \log \rho (x)$ of the stationary state-probabilities.

\subsection{Equilibrium case}\label{eqc}
The basic idea to define heat capacity $C(\beta)$ as in \eqref{hcao} is unchanged when moving between an equilibrium and a nonequilibrium situation, but obviously, equilibrium comes with a simplification.\\
To evaluate \eqref{qpt} in thermal equilibrium, we first note that then excess heat just means heat and $\dot{q}(\beta) = \dot{q}(\beta+\id \beta)=0$.  Furthermore, for reversible transformations when no work is done on the system, the First Law reads ${\dot{q}}_t(x, \beta+\id \beta) = - \frac{\id}{\id t} \langle E(x_t)\rangle_{\beta+\id \beta}$ for $x_0=x$,  in terms of the energy function $E$ of the system.  Hence, when there is no external (nonequilibrium) driving, \eqref{qpt} becomes $V(x,\beta+\id \beta) = E(x) - U(\beta+\id \beta)$, where $U(\beta)= \langle E(x)\rangle_{\beta} $ is the (average) equilibrium energy.  As a result, \eqref{hcao} then gives the well-known expression for the heat capacity at constant volume: $C(\beta)=-\beta^2\frac{\id }{\id \beta}U(\beta)$. That formula no longer holds in steady nonequilibrium where, even at constant volume, there is the possibility of irreversible work. \\
For comparison with the fluctuation formula \eqref{fluc}, we recall that when the system is in canonical equilibrium with partition function $Z$, 
\begin{eqnarray}
    \frac{\id \log \rho }{\id \beta}(x) &=& \beta^2\,E(x) +\beta^2\frac{\id \log Z}{\id \beta}\nonumber\\
    V(x,\beta) &=& E(x) - U(\beta)
\end{eqnarray}
so that \eqref{fluc} reproduces then the equilibrium expression
\[
C(\beta) = \beta^2\,\big\langle \left(E- U(\beta)\right)^2\big\rangle_\beta  \geq 0
\]
in terms of the energy variance. However. the nonequilibrium heat capacity and hence the correlation \eqref{fluc} can get negative, as we will illustrate via Example \ref{heac}.


\section{Markov jump processes}\label{mjp}  
To understand the low-temperature properties of the heat capacity defined above, we need to add a class of models that imitate some of the main aspects of the quantum world.   The present section presents the setup and Section \ref{qai} gives extra explanation and interpretation.  In particular, the (open) systems that we model should be thought of as consisting of discrete degrees of freedom, where the zero-temperature condition is well-separated from its excitations.\\   Furthermore, because we explore the nonequilibrium regime, we take a dynamics that violates the condition of detailed balance.

\subsection{Setup}\label{set}
When reactions occur far from equilibrium, modeling with Markov jump processes becomes very relevant to understand their calorimetry, \cite{dm,kkm}.  Additionally, there is a large literature on chemical reaction networks with papers starting from exactly the same setup, \cite{lux1,lux2,lux3,lux4,viv,sch}.
Such Markov dynamics with discrete states can be obtained as the weak-coupling limit of a quantum system, which is in incoherent contact with one or several heat baths; see more in Section \ref{qai}.  We refer to \cite{kamp} for more context on stochastic processes in physics and chemistry.

\begin{figure}[H]
\centering
\includegraphics[scale=0.9]{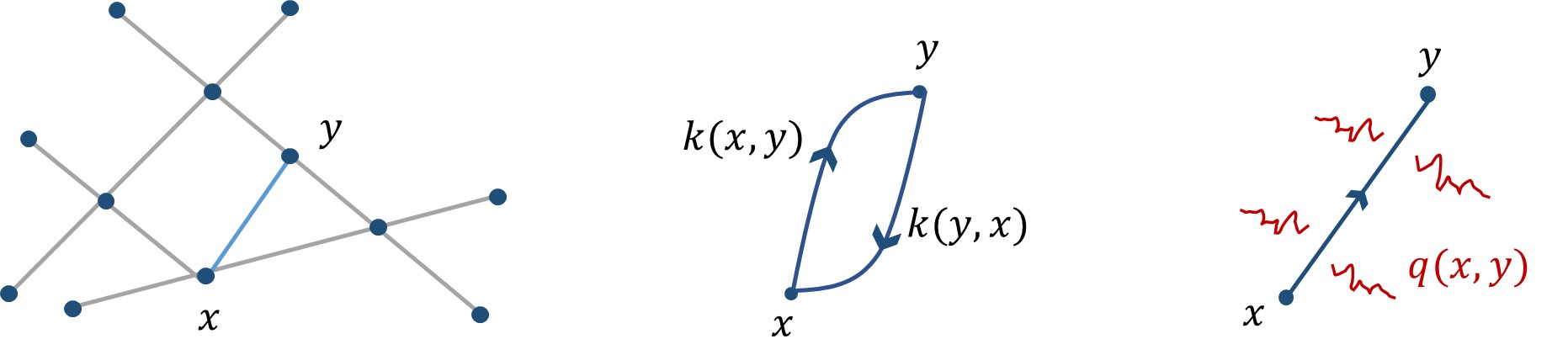}
\caption{\small{left: graph with vertices $x$ and $y$, connected via an edge. middle: the transition rates for the directed jumps over the edge. right:  indication of heat given or taken during the transition between the states $x\rightarrow y$; see \eqref{ldb}.}}\label{rate1}
\end{figure}

For a formal description, we consider Markov jump processes on an (arbitrary) finite, simple, and connected graph $\cal G$.  Vertices $x,y,z,\ldots$ denote the physical states or conditions on a mesoscopic scale of description. We emphasize that the states $x, y,\dots$ may represent many-body configurations such as indicating the quantity of a certain substance.  The edges indicate the possible chemomechanical transitions between configurations, including discrete changes in position or energy.  
We denote the transition rates by $k(x,y)>0$ for a jump $x\rightarrow y$, and  $k(y,x)>0$ is the rate for a jump $y\rightarrow x$. See Fig.~\ref{rate1} for a summary.\\

The transition rates determine a unique stationary distribution $\rho (x)>0$, solution of the stationary Master equation: for all states $x$,
\begin{equation}\label{std} \sum_y \big[k(y,x)\rho (y) - k(x,y)\rho (x)\big] =0
\end{equation}
We also introduce the backward generator $L$  of the Markov process, satisfying
\begin{eqnarray}\label{backw}
Lh \,(x) &=& \sum_y k(x,y)[h(y)-h(x)]\\
e^{tL}h\,(x) &=& \langle h(X_t)\,|\,X_0=x\rangle\label{etl}
\end{eqnarray}
for the expected value at time $t\geq 0$ of a function $h$,  when starting the process at $X_0=x.$  We refer to textbooks like \cite{kamp,grim}  for more introduction to Markov jump processes.\\

The rates $k(x,y)$  contain kinetic and thermodynamic information summarizing the weak coupling of the system to a heat bath at inverse temperature $\beta $. The rates may depend on 
 external parameters but, for simplicity, we concentrate here on the dependency of the rates on temperature (often as a subscript in $k_\beta(x,y)$).\\

\subsection{Quasipotential and heat capacity}
The crucial link between the mathematics of jump processes and their thermodynamic interpretation is contained in  the condition of local detailed balance \cite{ldb, Kat, les},
\begin{equation}\label{ldb}
\log\frac{k_\beta(x,y)}{k_\beta(y,x)} = \beta\, q(x,y)
\end{equation}
where  $q(x,y)=-q(y,x)$ is required to be the heat dissipated in the heat bath at inverse temperature $\beta$; see \cite{ldb} and references therein (and Fig.~\ref{rate1} for a graphical reminder).\\  
The {\it local} detailed balance has nothing to do with the notion of local equilibrium.  Rather, it must be contrasted with {\it global} detailed balance, where {\it all} currents vanish. When there is global detailed balance (which we take as defining equilibrium dynamics), there is an energy state function $E$ for which that heat in \eqref{ldb} equals $q_\text{eq}(x,y) = E(x) - E(y)$, for all edges $(x,y)$.\\
We also note that local detailed balance should be formulated for each dissipative transition channel separately, and it is easily broken by taken further coarse-graining or combining rates; see e.g. \cite{ast}.


Recall \eqref{qpt}  and how the quasipotential $V(x,\beta)$ appears in the expression \eqref{hcao} of the nonequilibrium heat capacity.  From now on, we write $V(x) = V(x, \beta) $ if no confusion with temperatures can arise.  Then, using the above formalism we have
\begin{equation}\label{qpv}
    V(x) = \int_0^\infty\Big[ \big\langle \dot{q}(X_t)\,|\,X_0=x\big\rangle - \dot{q}_\beta    \Big]
\end{equation}
where
\begin{eqnarray}\label{ju}
\dot{q}(x) &=& \sum_y k_{\beta}(x,y) \, q(x,y)\\
\dot{q} _\beta &=& \sum_x \rho _\beta(x)\,\dot{q}(x)
\end{eqnarray}
and $\rho _\beta(x)$ is the stationary distribution, solution of \eqref{std}. The above formul{\ae} realize the previous \eqref{qpt}.  In other words, the quasipotential \eqref{qpv} must be used in the heat capacity \eqref{hcao}, 
\begin{equation}\label{hca}
C(\beta) =  \beta^2 \, \big\langle \frac{\id V}{\id \beta}  \big\rangle_\beta 
\end{equation}
and, similarly, in the fluctuation formula \eqref{fluc}  for the correlation function
\begin{equation}\label{hcaf}
C(\beta) =  -\beta^2 \, \big\langle \frac{\id \log\rho }{\id \beta}(x)\, V(x)  \big\rangle_\beta 
\end{equation}
Note here that the stationary distribution $\rho$, solution of \eqref{std}, is in general not explicitly known.

\subsection{Example: molecular switch}\label{heac}
When a molecule can reversibly switch between two stable states, we have a molecular two-level switch.  It has various realizations, see e.g. \cite{ms,ms2}, and can be engineered according to various protocols of switching.  Here we take a toy-version, without being specific about the molecule, but focusing on the essential nonequilibrium character and its thermal response.\\

\begin{figure}[H]
			\includegraphics[scale=0.8]{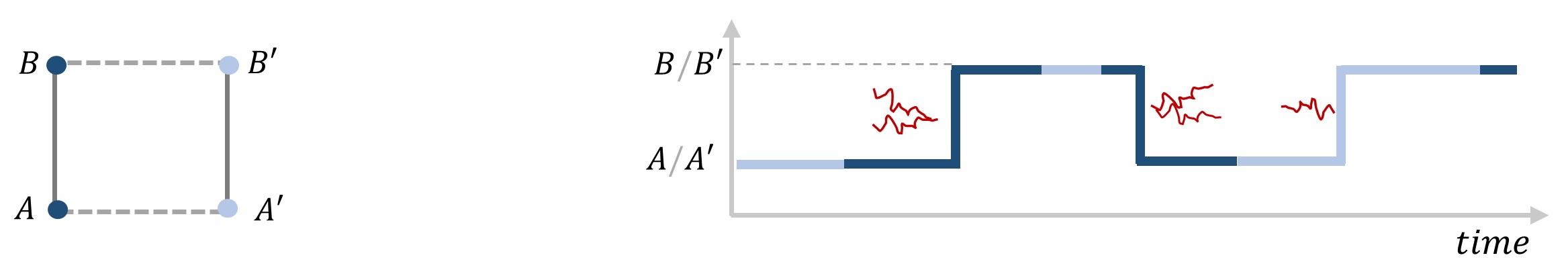}  
		\caption{\small{Molecular switch as described in \eqref{qs}. left: The possible transitions between the four states.  States $A-B$ and states $A'-B'$ are separated by an energy $\ve$. The horizontal transitions correspond to the switches.  right: A sample trajectory with the switching indicated by the change in color.  The heat is suggested as in Fig.~\ref{rate1}(right) when the system jumps between energy levels.}}
	\label{qsw}
\end{figure}

Consider the reaction scheme $A \rightleftarrows B$ and $A' \rightleftarrows B'$, where each time we imagine two molecular levels separated by energy $\ve$, but where, by a random environmental stimulus, shifts $A\leftrightarrow A'$  and $B\leftrightarrow B'$ occur at a certain rate $\alpha$; see Fig.~\ref{qsw}. 

\begin{figure}[H]
	\begin{subfigure}{(a)}
		\includegraphics[scale=0.58]{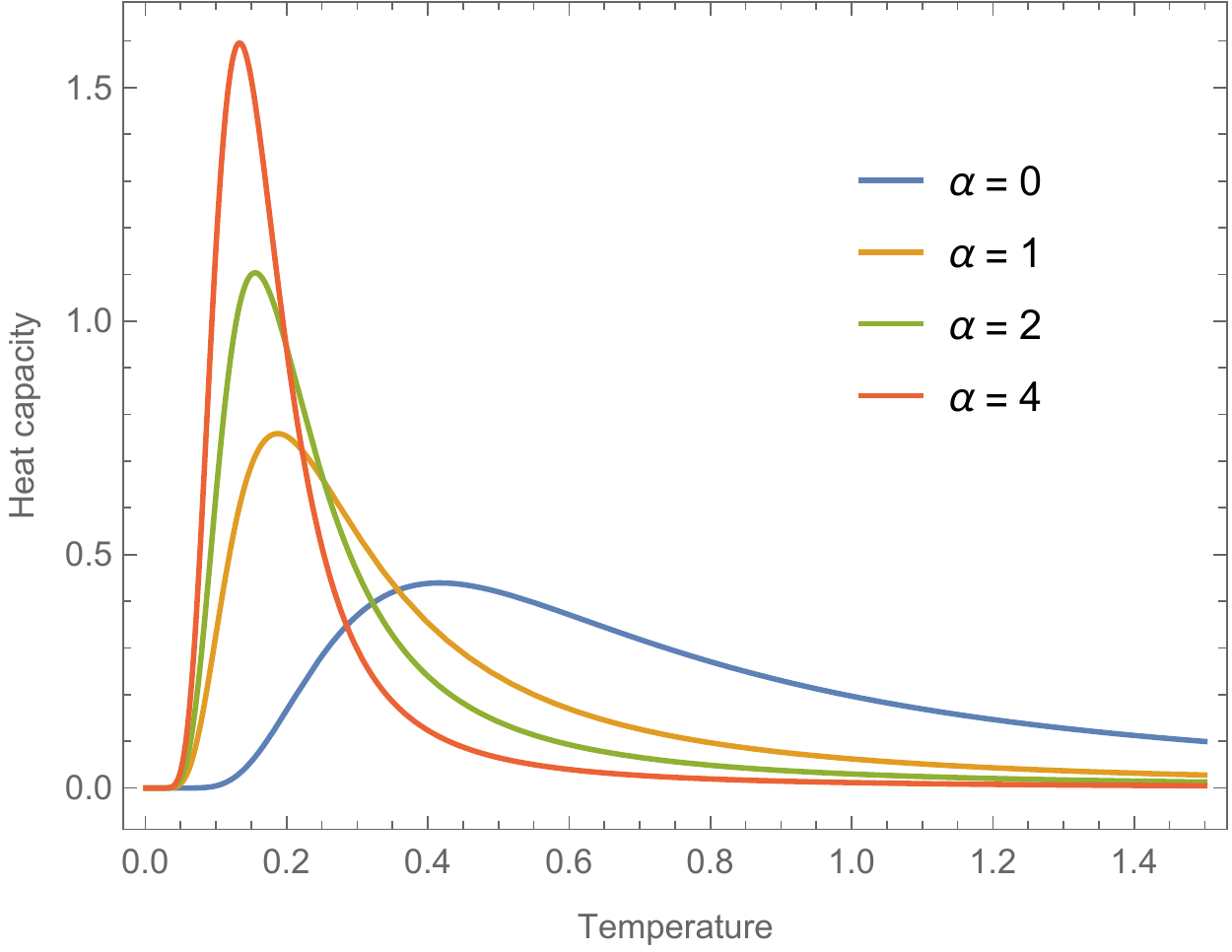}  
	\end{subfigure}
	\begin{subfigure}{(b)}
		\includegraphics[scale=0.58]{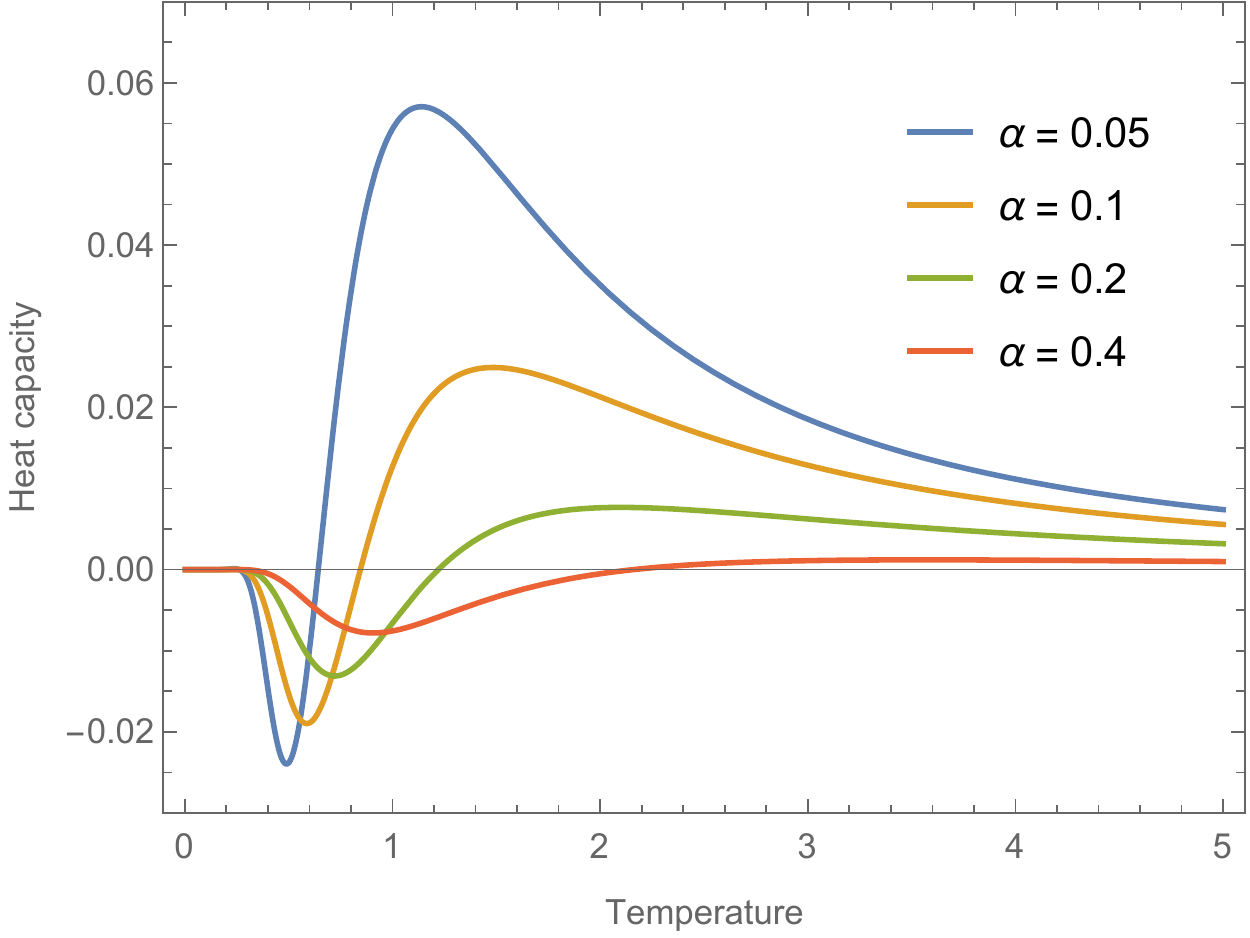}  
	\end{subfigure}
    \caption{\small{Heat capacity for the two-level switch defined in \eqref{qs}, (a) without kinetic barrier, (b) with  a barrier at $\Delta = 2$.}}
	\label{arrhenius}
\end{figure}
We thus distinguish four states $x \in \{A,A',B,B'\} $.  The transitions $A\leftrightarrow A'$ and $B\leftrightarrow B'$ occur at fixed rate  $k(A,A') = k(A',A) = \al = k(B',B) = k(B,B')$ for $\alpha \geq 0$.  The other four transitions are thermal at inverse temperature $\beta$ with rates
\begin{eqnarray}
	k(A,B) = k(B',A') &=& \nu\,e^{\be\varepsilon/2}\label{qs}\\
	k(A',B') = k(B,A) &=& \nu\, e^{-\be\varepsilon/2}\nonumber
\end{eqnarray}
The kinetic factor 
$\nu>0$ is a characteristic frequency for transitions between both levels. Detailed balance is broken for $\alpha\ve\neq 0$ since there is a difference between the product of rates along two opposite orientations of a loop: the orientation
$A\rightarrow B \rightarrow B' \rightarrow A' \rightarrow A$ has $ \al^2 \nu^{2} e^{\be\varepsilon}$ as product of its rates, whereas 
the reversed order makes the product $ \al^2 \nu^{2} e^{-\be\varepsilon}$.\\

The heat capacity \eqref{hca} can be calculated exactly along the lines of the previous subsection.
Fig.~\ref{arrhenius}(left) gives the heat capacity as a function of temperature for $\ve=\nu = 1$.  Observe that  whenever $\al \neq 0$ the heat capacity becomes massively enhanced at (very) low temperatures.
In Fig.~\ref{arrhenius} (right), we take $\nu = e^{-\be\Delta}$ which strongly depends on temperature with kinetic barrier $\Delta>0$. In Fig.~\ref{arrhenius} (right), we use $\Delta=2$.  Interestingly, the heat capacity takes both positive and negative values.   The low-temperature behavior is very different both from the (nonequilibrium) heat-bath dynamics in Fig.~\ref{arrhenius}(left) and from the equilibrium case, while still $C(T)\rightarrow 0$ as the temperature reaches absolute zero.

\section{Extended Third Law for Nonequilibria}\label{nerp}
Even for equilibrium system, the Third Law of Thermodynamics comes with a condition: the ground state must be nondegenerate to avoid residual entropy. For the extension of the present paper we need a similar (and first) static condition, to be discussed in the following subsection.  However, for our nonequilibrium purpose we also have a (second) dynamic condition which is the subject of Section \ref{proofs}. The main issue then for an extended Nernst Postulate is to explain  when and how the nonequilibrium heat capacity \eqref{hca} vanishes with temperature.   More specifically, for \eqref{hca}--\eqref{hcaf}, to have
\begin{equation}\label{3rd}
C(\beta) \rightarrow 0, \;\text{ with }\, \beta\uparrow \infty    
\end{equation}
The arguments make the contents of the following two subsections.  We focus here on some of the essential ingredients; mathematical details are in \cite{arxiv}.\\
All considerations about thermal properties of low-temperature processes must be compatible with \eqref{3rd} (when the static and dynamic criterion are satisfied).   The illustrations of Section \ref{exs} show how the conditions are exactly on target.   

\subsection{Low--temperature asymptotics of static fluctuations}
From Section \ref{set}, there is a unique stationary probability distribution $\rho_\beta>0$ satisfying \eqref{std}.
At low temperatures, there exists an approximate expression for $\rho_\beta(x)$.
Not too surprisingly, it is determined by the asymptotics of the transition rates,
\begin{equation}\label{fifi}
\phi(x,y) = \lim_\beta \frac 1{\beta} \log k_\beta(x,y)
\end{equation}
From an analysis of the Kirchhoff formula in \cite{heatb,lowT,intr}, we have found that for $\beta\uparrow \infty$,
\begin{equation}\label{lowkir}
    \rho _\beta(x)\simeq e^{-\beta [\phi^*-\phi(x)]}
\end{equation}
with 
\begin{equation}\label{phi}
\phi^* =\max_z \phi(z), \quad \phi(z) = \max_{\cal T} \sum_{(x,y) \in \cal T_z} \phi(x,y)
\end{equation}
where the maximum is over all spanning trees $\cal T$ on the graph, and with $\cal T_z$ the oriented tree obtained from $\cal T$ by directing it toward $z$.  Any state $x^*$ satisfying $\phi^* =\phi(x^*)$ is called dominant; they have the largest probability in the large-$\beta$ asymptotics.  We think of dominant states as the nonequilibrium analogue of equilibrium ground states, but note that they do not need to minimize energy.  The asymptotics \eqref{lowkir} has to be understood in the logarithmic sense, meaning that there may be a prefactor, depending on $x$, which is not exponentially small in $\beta$.\\

Actually, \eqref{lowkir} is a general result for the Markov jump processes we consider here, but for the present purpose of establishing an extended Third Law, it suffices that there is a unique state $x^*$ so that
\begin{equation}\label{c22}
        \rho_\beta (x)=\delta_{x,x^*}+ O(\beta^{-2})
\end{equation}
as our first criterion.  It means that $   \rho_\beta (x) \rightarrow 0$ for $x\neq x^*$, decays to zero with temperature at least as fast as $T^2$.  It implies $\frac{\id}{\id \beta}\, \rho_\beta^\text{s}\rightarrow 0$ as $\beta\uparrow \infty$.  Checking with \eqref{hcaf}, the behavior \eqref{c22} will send the heat capacity to zero when the quasipotential $V_\beta$ remains bounded in the limit $\beta\uparrow \infty$.  Requiring the behavior \eqref{c22} is therefore a first (static) condition ensuring the validity of the extended Third Law.\\ 

Note also that the decay rate     $\rho_\beta (x) \rightarrow 0$ for $x\neq x^*$ reveals the support of the zero-temperature distribution, much as in equilibrium.  See \eqref{lowkir}:  if $x^*$ is the unique dominant state, i.e., the unique maximizer of $\phi(z)$ in the sense that all other states are suppressed by $O(e^{-\tilde{\ve} \beta}$) for some $\tilde{\ve}>0$, then the convergence in \eqref{c22}  is actually exponentially fast.  If there are multiple dominant states (i.e., maximizers of $\phi(z)$), then the convergence $\id \rho _\beta \rightarrow 0$ may be polynomial in temperature. 
 As a result, the heat capacity goes to zero exponentially fast in $\beta\uparrow \infty$ in the case of a unique dominant state and possibly much slower due to the presence of  more than one dominant state.\\
All that is shared with the equilibrium situation.  There is however another and second condition which is dynamic, and addresses the boundedness of $V_\beta$ (which is never an issue in equilibrium; see Section \ref{eqc}).

\subsection{Boundedness of the quasipotential}\label{proofs}

We come to the statement and explanation of a second  and dynamical condition for the nonequilibrium version of the Third Law: a sufficient criterion for the quasipotential $V$ to be uniformly bounded at vanishing temperature. \\
 As is visible from \eqref{qpv}, boundedness of $V$ as a function of $\beta\uparrow\infty$ strongly relates with low-temperature relaxation properties and in particular with the accessibility of the states that support the zero-temperature condition; we will see that explicitly in  Section \ref{exs}. \\

For the actual argument, which is different from the graphical arguments in \cite{arxiv}, we start with a remark.  If for all edges $(xy)$ in the graph, the differences $|V(x,\beta) - V(y,\beta)|$ remain  bounded, then also the quasipotential $V$ must remain bounded, because, under our conditions, $V(x^*,\beta)$ cannot be diverging.\\ 
Let us, therefore, consider differences of \eqref{qpv},
\begin{equation}\label{dprv}
    V(x,\beta) - V(y,\beta) = \int_0^{+\infty} \id t\,
    [\,\langle {\dot{q}}(X_t)\,|\,X_0=x\rangle - \langle {\dot{q}}(Y_t)\,|\,X_0=y\rangle\,]
\end{equation}
where we indicated with $X_t$, respectively $Y_t$, random walkers starting from initial position $x$, respectively $y$.  We couple those two walkers, i.e., we consider the process of coupled random walkers $(X_t,Y_t)$ with $(X_0,Y_0)=(x,y)$ on the double graph.  In an optimal coupling, once they meet for the first time, $X_t=Y_t$ at a random time $t=t_M(x,x';\beta)$, they keep on taking the same path.  Surely, for every finite $\beta$ there is such a first time $t_M(x,y;\beta)$, and $X_t=Y_t$ for all $t\geq t_M(x,y;\beta)$.  Let $\tau_M = \langle t_M(x,y;\beta)\rangle$ be the average first meeting time. Then, an upper bound on the integral \eqref{dprv} is given by
\begin{equation}\label{dy}
|V_\beta(x) - V_\beta(y)| \leq R\,\tau_M \,\max_{x',y'} |{\dot{q}}(x')-{\dot{q}}(y')|
\end{equation}
where $R$ is the linear size of the graph (maximal graph distance between any two vertices). The time $\tau_M$ is bounded by the relaxation time  $\tau$.  On the other hand, the expected difference in the power $\dot{q}$, while fluctuating, can be estimated, at least far away from equilibrium,  by a quantity proportional to the dissipation rate $\tau_d^{-1}$. The latter refers to the dissipation time $\tau_d^{-1}$, estimated as the mean power divided by the heat. Therefore, to have \eqref{dy} bounded, it suffices that relaxation times $\tau$ do not exceed the dissipation time $\tau_d$:
\begin{equation}\label{rd}
\tau < \tau_d
\end{equation}
Or, our second condition on the transition rates states that the mutual differences in relaxation times do not exceed the dissipation time scale.  In the case of detailed balance, the dissipation time $\tau_d = \infty$ and $V_\beta(x) = E(x) -\langle E\rangle_\beta$ is automatically bounded indeed.\\

For a different yet manageable mathematical approach, we  observe that \eqref{qpv} is  of the form
\[
V(x,\beta) = \int_0^\infty\id t \,(e^{tL} \dot{q}(x) - \dot q_\beta )
\]
where we used \eqref{etl} with the backward generator $L$. Doing the integral over time $t$, we see that \eqref{qpv} is  equivalent with
\[
LV\,(x,\beta)=  \dot q_\beta 
- \dot{q}(x)
\] 
Then, using \eqref{backw} and substituting \eqref{ju}, we get the linear equations
\begin{equation}\label{eg}
\sum_{y}k_\beta(x,y)[V(y,\beta) - V(x,\beta) + q(x,y)] =  \dot q_\beta 
\end{equation}
to be satisfied for all states $x$ (while the right-hand side does not depend on $x$).  
The solution is in terms of a graphical representation of the $V(x,\beta) - V(y,\beta)$; see \cite{arxiv} where the main step is an application of the matrix-forest theorem.  
To understand the issue, note that when, in the limit $\beta\uparrow\infty$, the transition rates $k_\beta(x,y) \sim \exp (-c \beta)$ for some $c>0$, get exponentially smaller than the dissipated power $\dot q_\beta $, the differences $|V(x,\beta)-V(y,\beta)|$ may grow exponentially with $\beta$ and yet satisfy \eqref{eg}.  Since we want bounded $V$, that needs to be avoided.  That will be illustrated in Example \ref{e2}.\\

For getting more concrete, suppose that in \eqref{fifi}
\begin{eqnarray}\label{lb}
\phi(x,y) &=& 0 \quad \qquad \text{ when }q(x,y)\; \geq 0\\
&=& q(x,y) \;\,\text{ when }  q(x,y) < 0\nonumber
\end{eqnarray}
in terms of the heat $q(x,y)$ obtained in \eqref{ldb}, so that $\phi^*\leq 0$.  The $\phi^*$ was introduced in \eqref{phi} and is dominated by the accessibility of the most probable state $x^*$; see \eqref{c22}.   An extension of the Kirchhoff formula applied to \eqref{eg} (the matrix-forest theorem as explained in \cite{arxiv}), shows that asymptotically
$V(x,\beta)= v_\beta(x)\,e^{-\beta \phi^*}$ where $v_\beta(x)$ is uniformly bounded in $\beta\uparrow \infty$.  Therefore, to have $V(x,\beta)$ bounded, it suffices that $\phi^* = 0$ in \eqref{phi}.  That is the case when there is a spanning tree with edges directed towards $x^*$ along which all $q(x,y) \geq 0$.  Hence,  it certainly suffices that the digraph, the directed subgraph obtained by only keeping the edges where $q(x,y) \geq 0$, is sufficiently well-connected to allow reaching $x^*$ from everywhere in the graph.\\We conclude that for the extended Third Law to hold, in addition to \eqref{c22}, it suffices there is a rooted spanning tree in the graph along edges where the dissipated heat $q(x,y) \geq 0$. Then, the heat capacity vanishes at absolute zero, as announced in \eqref{3rd}.\\
 As an extreme example, on the complete graph and for every choice of rates realizing \eqref{lb}, the extended Third Law always holds, because then every edge has a direction where $q(x,y) \geq 0$, and we know that a tournament has a Hamiltonian path; see \cite{book}.  The fact that for denser graphs the extended Third Law gets satisfied easier (because of greater accessibility of $x^*$), is of course compatible with having smaller relaxation times, as we used before.  It also indicates the role of zero-temperature tunneling: as it increases the connectivity of the graph, it promotes the validity of the extended Third Law.

\section{Illustrations}\label{exs}
The following  examples explain in more concrete ways how to think of the (extra with respect to equilibrium) dynamical condition for the nonequilibrium Third Law.  Counterexamples allow us to see in what sense that condition is natural, hence necessary in some sense.

\subsection{Driven reaction with source/sink}\label{e1}
Consider the three-cycle $A\leftrightarrow B\leftrightarrow C \leftrightarrow A$, while all three states are connected to $Z$.

\begin{figure}[H]
\includegraphics[scale=0.8]{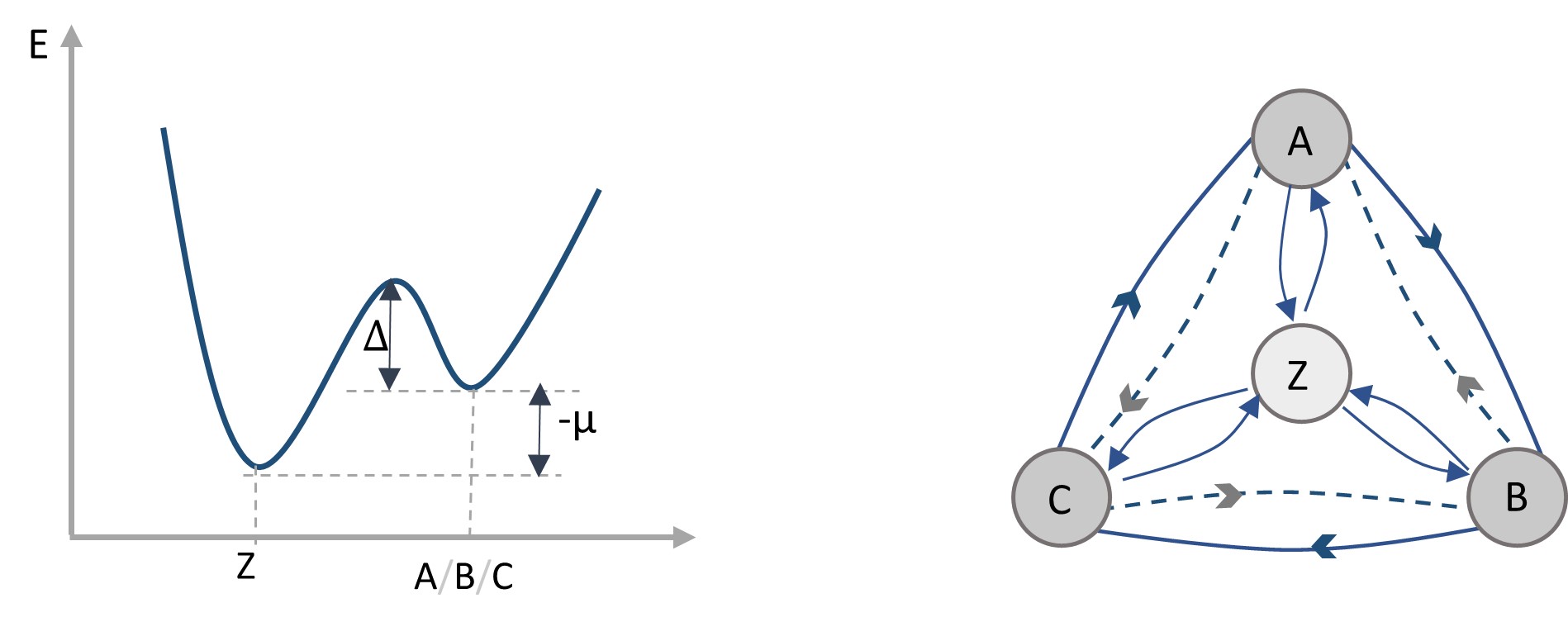}
\caption{\small{right: Cartoon depicting a circuit of three states $A\leftrightarrow B\leftrightarrow C \leftrightarrow A$, and  possible transitions to a sink/source state $Z$ in the center. left: the energy landscape modeling the transitions between the states in the cycle and the source/sink.}}\label{arr}
\end{figure}
The transition rates are
\begin{eqnarray}\label{hop1}
        k(A,B)=k(B,C)=k(C,A)&=&\frac{\nu}{1+e^{-\beta\zeta}}\;\nonumber\\
k(B,A) =k(C,B)=k(A,C)        &=& \frac{\nu}{1+e^{\beta\zeta}}\;\text{ when } \eta_{i+1} = 1 = 1-\eta_{i}\nonumber
    \end{eqnarray}
where  $\zeta> 0$ is the work done by a driving force, making $A\rightarrow B \rightarrow C\rightarrow A$ a dissipative cycle.\\  Furthermore, there are transitions from and toward a sink/source state $Z$, with
\begin{eqnarray}\label{exen}
k(A,Z)=k(B,Z) = k(C,Z)&=& \tilde{\nu}e^{-\beta \Delta}\nonumber\\
k(Z,A)=k(Z,B) = k(Z,C)&=&\tilde{\nu}\,e^{-\beta \Delta} e^{\beta\mu}
\end{eqnarray}
where $\mu$ plays the role of a chemical potential, and $\Delta\geq 0$ is a kinetic barrier for entering of leaving the dissipative cycle.
 for loading a dot.  Here, $\mu$ is the chemical potential at zero temperature (i.e., the Fermi energy) and the energy $a\geq 0$ is a kinetic barrier.\\
Using the above scheme, 
the heat capacity \eqref{hca}  has been calculated (see \cite{pnas} for a similar case) and was found to be,
\begin{equation}\label{cb}
\,C(\beta) = \frac{\beta^2\mu^2 e^{-\beta\mu}}{(1 +e^{-\beta\mu})^2}
-\beta^2 \mu\zeta\frac{\nu}{\tilde{\nu}}\,\frac{ (e^{\beta(a-3\mu)}-e^{\beta(a-2\mu)})}{(1+ e^{-\beta\mu})^4}\tanh(\beta\zeta/2)\
\end{equation}
The second term gathers the nonequilibrium (driving $\zeta$) and kinetic ($\Delta,\nu,\tilde{\nu}$) parameters.\\

We have Third Law  behavior when $\mu < -\Delta$ or $\mu > \Delta/2$; then $C(\beta\uparrow \infty)\to 0$.  Yet, small  $|\mu|/\Delta$ leads to violations: $C(\beta\uparrow \infty) \to \infty$ diverges for $-\Delta< \mu < \Delta/2$ when $\mu\,\Delta\,\zeta\,\nu \neq 0$, \cite{pnas}.
The previous section and  condition \eqref{rd} more specifically, make clear what happens: relaxation times $\tau$ get too large compared to the dissipation time $\tau_d$.  E.g., for $\mu<0$, the state $Z$ has the largest stationary probability, but the cyclic reaction typically runs for a time $\tau \propto \tilde{\nu}^{-1}\,e^{\beta \Delta}$ before the system enters state $Z$.  On the other hand, the dissipation time $\tau_d$ is given by the inverse of the current, $\tau_d\propto \nu^{-1}\, e^{-\mu\beta}$.  In other words, $\tau < \tau_d$ when $\Delta<-\mu$.  
Similarly, for $\mu >0$, the dissipation time is $\tau_d \propto  \nu^{-1}\, e^{\mu\beta}$  and the relaxation time is $\tau \propto \tilde{\nu}^{-1}\,e^{\beta (\Delta-\mu)}$ for the empty state (= to get stationary). Hence, the extended Third Law holds when $\tau < \tau_d$ or $\mu> \Delta/2$, but at a chemical potential $\mu=-\Delta$ and at $\mu=\Delta/2$ zero-temperature transitions in thermal properties occur, unseen in equilibrium $\zeta=0$.\\

\subsection{Avoiding delay --- staying connected}\label{e2}
\begin{figure}[H]
\includegraphics[scale=1.2]{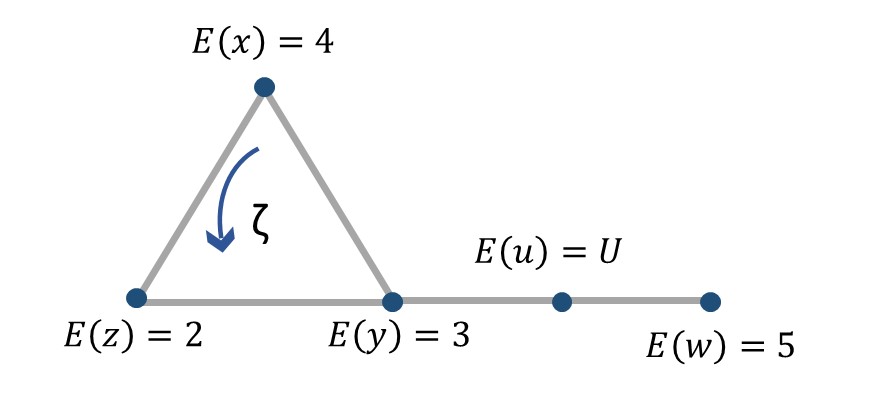}
\caption{\small{ A driven system with 5 states. Depending on $E(u)=U$ the quasipotential $V_\beta$ remains uniformly bounded ($U\leq 6$).}}\label{badhair}
\end{figure}

We illustrate the physics explained below \eqref{eg} with a more abstract example.\\ 
Consider the tree-loop of Fig.~\ref{badhair}, where also the constant driving of strength $\zeta$ and the energy landscape are indicated.  The driving is counter-clockwise for $\zeta>0$.
The dissipative structure is the triangle  $x\rightarrow z \rightarrow y \rightarrow x$ carrying a nonzero steady current. It is easy to verify that $z=x^*$ (see \eqref{lowkir}) is the unique dominant state and that the dissipative power $\dot{q}_\beta$ is of order $\exp(-\beta)$, which is the escape rate from $z$.  The energy difference over the outer edge $(uw)$ is $E(w) - E(u) = 5-U$, which means that $k_\beta(w,u) \simeq \exp[-\beta(U-5)]$ for $U>5$. Hence, in \eqref{eg}, we can take a bounded edge difference $V(w)-V(u)$ as long as $U\leq 6$.  There is a trap over the edge $(w,u)$ when $U$ is too large, with a too long delay for relaxation for the quasipotential in $V(w)$ to remain bounded.  That stands for the general scenario.  If we would add for example the possibility of tunneling between the vertices $x\leftrightarrow w$, the extended Third Law \eqref{3rd} would be satisfied again.\\  

\section{Quantum context}\label{qai}
The models we are using in the previous sections (Markov jump processes), concern systems without quantum coherence or, equivalently, in the weak coupling limit.  Alternatively, we can consider it as the dynamics one obtains in the Born-Oppenheimer approximation or via the Dirac-Fermi Golden Rule beyond the decoherence time.  In that sense, it is easiest to imagine a microscopic system for which we assume, for simplicity, that the decoherence time is much smaller than the relaxation time.\\
In that same spirit of weak coupling, the dynamics we consider are Markovian, i.e. memoryless for the residence times.  That is why the formalism can be specific in its condition \eqref{ldb} of (local) detailed balance in terms of the transition rates.\\
Note also that diffusion processes, as modeled by Langevin or Smoluchowski equations, are no good for discussing very low temperatures in the same way as classical (Newtonian) physics breaks down there.  In particular, for exactly the same reason as in equilibrium, we cannot expect an extended Third Law for (nonequilibrium) diffusion processes.\\

Let us finally mention a point that we learn from the previous examples illustrating the issue of obtaining the boundedness of the quasipotential.  Interestingly, the validity of the extended Third Law gets enhanced when the low-temperature process does not get trapped or frozen in certain states that do not contribute to the stationary dissipative condition.  In other words, dynamic activity helps. That then is a feature that is typically quantum as we expect that due to quantum fluctuations, the dynamics remains sufficiently active even at close-to-zero temperature, e.g. via tunneling events or, even more generally, just because of the Heisenberg uncertainty relation.

\section{Significance}\label{sig}
  Extending the Third Law to nonequilibrium processes, delivers a nontrivial result on thermal properties of steady states.  As a matter of fact, very few such general results exist at all.
  The main operational significance of the extended Third Law is that heat capacities become zero when moving to very low temperatures, universally in all parameters.
  We emphasize that the operational and physically-strongest meaning of the Third Law, both in equilibrium and in nonequilibrium, does not need to proceed via the Second Law at all.
Heat, excess heat, and heat capacity are all perfectly well-defined and have physical meaning without the existence of entropy.  
Obviously, the Second Law is satisfied (independently) and there is strict positivity of the entropy production, but it plays no role here and that is normal.  Even in equilibrium, entropies are measured via heat capacity, and the Third Law gets experimentally validated by low-temperature calorimetry.
As the standard (equilibrium) Third Law but now extended to nonequilibria, that sets constraints to all low-temperature transformations.  Even though the total net heat flux cannot vanish now, the heat capacity  indeed does go to zero for a large class of nonequilibrium systems.\\

While the extended Third Law may well be an important result on the thermal susceptibility of nonequilibrium systems at low ambient temperature, measuring nonequilibrium heat capacities does constitute a new challenge, \cite{dio,cera,kkm}.  The extended Third Law begs for experimental confirmation.
As explained in \cite{calo,dm,ND,pnas}, an AC-calorimetric  scheme should work perfectly well and we hope the present work motivates further experimental work on low-temperature calorimetry for out-of-equilibrium systems.

\section{Conclusion}\label{con}
Under a static ``nondegeneracy'' condition and a dynamic ``no-trapping'' condition, we have argued that  the steady nonequilibrium heat capacity tends to zero with temperature.  Both general theoretical arguments and specific examples have been used to specify those conditions.\\

Concerning the static condition, exponential {\it versus} polynomial decay of the heat capacity, when the ambient temperature tends to zero, informs about the degeneracy structure of nonequilibrium steady systems.  The dependency of the low-temperature heat capacity on the behavior of the stationary condition $\rho _\beta$ is an important instance of how heat, also in nonequilibrium statistical mechanics, informs about low-temperature degeneracy, extending the connection which exists in equilibrium between the Clausius heat and the Boltzmann entropy, \cite{Balianbookvolume1}.\\
Interestingly, our nonequilibrium extension also necessitates an (extra) dynamical condition, which asks for sufficient connectivity of the graph of states, allowing low-temperature dynamical activity so that relaxation times are smaller than the dissipation time.\\

\noindent {\bf Acknowledgments:}  We are grateful to Pritha Dolai for discussions on the example in Fig.~\ref{arr}, see \cite{pnas}, and to Irene Maes for discussions on some mathematical aspects, see \cite{arxiv}.\\

\noindent{\bf Conflict of Interest:} The authors  have no conflicts to disclose.\\
\noindent{\bf Data Availability Statement:} The data that supports the findings of this study are available within the article .\\

\end{document}